\documentclass[aps,pre,twocolumn,showpacs,superscriptaddress,amsmath,amssymb]{revtex4-2}

\usepackage{graphicx}
\usepackage{amsmath}
\usepackage{bm}
\usepackage{color}

\begin{document}

\title{Extended self-similarity in two-dimensional complex plasmas}

\author{V. Nosenko}
\email{V.Nosenko@dlr.de}
\affiliation{Institut f\"{u}r Materialphysik im Weltraum, Deutsches Zentrum f\"{u}r Luft- und Raumfahrt (DLR), D-51170 Cologne, Germany}
\affiliation{Center for Astrophysics, Space Physics, and Engineering Research, Baylor University, Waco, Texas 76798-7310, USA}

\date{\today}
\begin{abstract}
Self-similarity is a property of an object or process where a part is similar to the whole. Mathematically, it can often be expressed as a power-law scaling of the quantity of interest. Extended self-similarity is a concept widely used in the field of turbulence and refers to the power-law scaling of the longitudinal structure functions of the velocity field expressed through the structure functions of different orders rather than distance. Originally discovered by Benzi {\it et al.} \cite{Benzi:1993} in a fully developed turbulence, it was later found to hold in other situations and systems. In this paper, we show that in an active-matter system, extended self-similarity is possible even without the presence of respective power-law scaling in the underlying structure functions of distance. The active-matter system used in this study was a single-layer suspension of active Janus particles in a plasma. Janus particles are polymer microspheres with hemispherical metal coating. When dispersed in a plasma, they acquire self-propulsion and act as microswimmers. Extended self-similarity was also observed in the velocity field of a single-layer suspension of laser-heated regular (passive) particles, where the underlying structure functions displayed a hint at the power-law scaling near the mean interparticle distance. Therefore, it appears to be an inherent characteristic of complex plasmas.
\end{abstract}


\maketitle

\section{Introduction}

Self-similarity, also known as scale-invariance, is a property of an object or process where a part is similar to the whole. Self-similarity is ubiquitous in nature, for example in coastline shapes \cite{Mandelbrot:1967}, sand wave dynamics \cite{Nikora:2001}, atmospheric boundary layer wind fields \cite{Kiliyanpilakkil:2015}, and energy spectrum in the inertial interval of fully developed turbulence \cite{Benzi:2023}. Mathematically, such lack of dependence of a certain value on spatial or temporal scales can be expressed, for example, via a power-law function $f(x)=Ax^\alpha$, which satisfies the homogeneity relation$f(\lambda x)=\lambda^\alpha f(x)$, where $\lambda$ is a scale factor and $\alpha$ is a scaling exponent.

Extended self-similarity (ESS) is a concept originally proposed by Benzi {\it et al.} \cite{Benzi:1993} for a fully developed turbulence. It was later found to hold in many other situations and systems from the physics of atmosphere \cite{Kiliyanpilakkil:2015} to geology \cite{Nikora:2001} and finance \cite{Constantin:2001}. ESS refers to the power-law scaling of the structure functions of the velocity field (or other suitable quantity) expressed through the structure functions of different orders rather than distance (or time). Extended self-similarity of the velocity fields in fluids is particularly important because it can help to identify scaling regimes otherwise obscured and can even provide a more accurate method of measuring the scaling exponents \cite{Benzi:2023}. Despite the success of ESS concept, it has not yet been fully explained theoretically \cite{Chakraborty:2010}. An explanation based on the reduction of subdominant contributions to scaling in the ESS representation was proposed in Ref.~\cite{Chakraborty:2010}. Multifractal interpretation of ESS was proposed in Ref.~\cite{Benzi:1996}. Building a rigorous theory of ESS will be facilitated by new insight into its applicability to further experimental systems and situations.

Complex, or dusty plasmas are suspensions of micron-size solid particles in a plasma, often in the sheath area above the lower electrode in a gas discharge \cite{Beckers:2023}. The particles are usually charged negatively because they collect more electrons than ions upon initial contact with plasma. They interact with each other via a screened Coulomb (Yukawa) potential and form strongly coupled structures with liquid or even solid ordering known as plasma crystals. Since the particles are suspended in a rarefied gas, their dynamics are not overdamped. In addition, the particles can be imaged individually using fast video cameras. Due to these properties, complex plasmas are excellent model systems to study generic processes in soft condensed matter at the level of individual particles and in real time \cite{New_book}. They were successfully used to study phase transitions \cite{Thomas:1996,Nosenko:2009,Melzer:2013}, transport phenomena \cite{Nunomura:2006,Nosenko:04PRL_visc,Hartmann:2011,Nosenko:08PRL_therm}, waves and instabilities \cite{Nunomura:2002,Piel:2002,Zhdanov:2003,Avinash:2003,Couedel:2010}, and turbulence \cite{Bajaj:2023}.

Active matter is a collection of active particles, each of which can extract energy from their environment and convert it into directed motion, thereby driving the whole system far from equilibrium \cite{Elgeti:2015,Bechinger:2016}. Janus particles have two halves of different properties, for example hemispherical metal coating on one side of a plastic particle. When suspended in a plasma, the Janus particles acquire self-propulsion and move in characteristic looped trajectories suggesting a combination of spinning and circling motion. The particle propulsion mechanism was identified as a combination of photophoretic force \cite{Du:2017,Wieben:2018,Nosenko:2020PRR_JP} and asymmetric ion-drag force \cite{Nosenko:2022}. The latter is similar to the force exerted on a non-spherical asymmetric particle from homogeneous, stationary, and isotropic plasma due to the scattering and collection of ions moving in the electric field of the asymmetric particle \cite{Krasheninnikov:2024}.

Using a complex plasma with active Janus particles as a model system allows one to study chaotic flows in a homogeneous isotropic system, which is stationary in time. In contrast, inertial (high-Reynolds-number) turbulence in a homogeneous isotropic system is necessarily declining with time without external drive \cite{Landau_v6}.

In this paper, a two-dimensional complex plasma with active Janus particles was studied experimentally. We show that the velocity field of the system of active Janus particles features extended self-similarity even though the underlying structure functions (of distance) lack respective power-law scaling. This finding is compared to the scaling in the laser-heated suspension of regular (passive) particles.

\section{Experimental method}

The Janus particles used in this study were prepared using the following method \cite{Nosenko:2020PRR_JP}. Melamine-formaldehyde (MF) microspheres \cite{microparticles} with a diameter of $9.27~\mu$m and mass $m=6.3\times10^{-13}$~kg were dispersed in isopropanol. A drop of the suspension was placed on a microscope cover glass. After isopropanol evaporated, the particles formed a monolayer on the glass surface. They were coated on one side with a $\approx 40$~nm layer of gold using a radio-frequency magnetron source. The resulting Janus particles were then collected by scratching them off the wafer by a sharp blade, placed in a standard container and dispensed into the plasma.

\begin{figure}[tbp!]
\centering
\includegraphics[width=1.0\columnwidth]{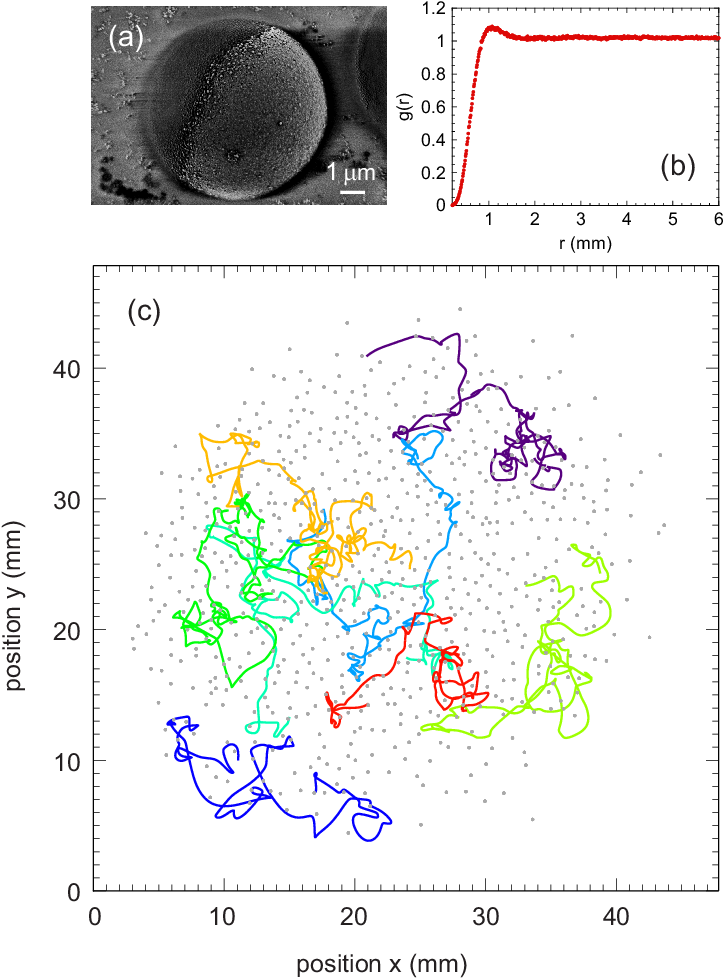}
\caption {\label {Fig1} (a) Scanning electron microscopy image of a Janus particle showing hemispherical coating with a $40$-nm layer of gold, (b) pair correlation function $g(r)$ of a single-layer suspension of Janus particles in plasma, (c) trajectories of eight selected Janus particles during $21.81$~s superimposed on the snapshot of all particle positions in the last frame of analysed video. The Janus particles were suspended as a single layer in rf plasma sheath. The argon pressure was $p_{\rm Ar}=0.66$~Pa, the rf discharge power was $P_{\rm rf}=20$~W, the illumination laser power was $P_{\rm laser}=14$~mW.
}
\end{figure}

We verified that the microparticles indeed received partial metal coating using scanning electron microscope (SEM) imaging. Coated particles were dispensed from the particle container onto a conductive substrate and imaged with the Zeiss Merlin electron microscope using the in-lens detector (annular scintillator detector in the column with optically coupled photomultiplier) of low-energy secondary electrons. The use of the in-lens detector allowed us to work at low primary electron energy ($1$~kV) and low beam current ($15$~pA) in order to be sensitive to thin metal layers and to avoid strong charging effects due to low electric conductivity of the microspheres. The SEM image in Fig.~\ref{Fig1}(a) shows a Janus particle with hemispherical gold coating containing small ($\lesssim70$~nm) inclusions, which are likely solidified droplets.

A single layer of the Janus particles was suspended in the plasma sheath above the lower electrode in a modified Gaseous Electronics Conference (GEC) radio-frequency (rf) reference cell \cite{Couedel:2022}. Plasma was produced by a capacitively coupled rf discharge in argon at $13.56$~MHz. The gas pressure was $0.66$~Pa, the discharge rf power was $20$~W.

The particle suspension was imaged from the top with the Photron FASTCAM mini WX100 video camera operating at $125$ frames per second (fps) and from the side with the Sony XC-ST50 camera operating at $30$ fps. The particles were illuminated by a horizontal laser sheet with the wavelengths of $660$~nm. In addition, a selected vertical cross-section of the particle suspension was illuminated by a vertical laser sheet with the wavelengths of $635$~nm. The cameras were equipped with matching bandpass interference filters.

% (b),(c) JP-13-4-L2.

\begin{figure}[tbp!]
\centering
\includegraphics[width=0.9\columnwidth]{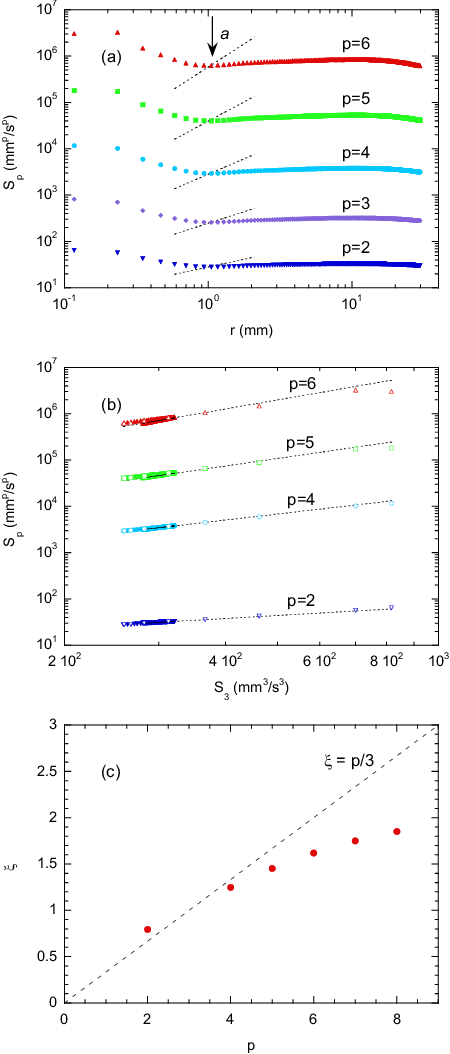}
\caption {\label {Fig2} (a) Longitudinal structure functions $S_p(r)$ for a single-layer suspension of Janus particles, (b) the same functions replotted as $S_p(S_3)$, (c) scaling exponent $\xi$ as a function of $p$. The dotted lines in panels (a) and (b) indicate the slopes defined by the Kolmogorovian exponents $\xi_p=p/3$. The power-law scaling of $S_p(S_3)$ reveals extended self-similarity of the velocity field. The arrow in panel (a) indicates the mean interparticle spacing $a$. The open symbols in panel (b) are for $r<a$, the solid symbols for $r\geq a$. The scaling exponents in panel (c) were obtained from the power-law fits of $S_p(S_3)$ (not shown here).
}
\end{figure}
% JP-13-4-L2.

The particle motion was recorded during $21.8$~s ($2726$ frames at $125$ fps). The video was analyzed in the following way. The particle positions were measured in every frame with subpixel resolution using a moment method \cite{SPIT}. The particles were traced from frame to frame and their velocities were calculated from the respective displacements. From the particle positions and velocities, various spatio-temporal correlation functions were calculated.

\section{Results and discussion}

\begin{figure}[tbp!]
\centering
\includegraphics[width=0.9\columnwidth]{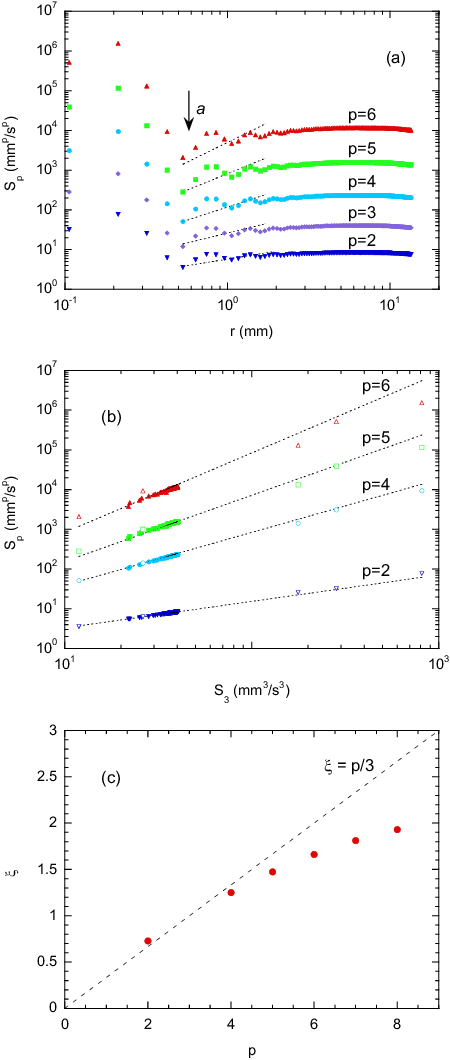}
\caption {\label {Fig3} Same as Fig.~\ref{Fig2}, but for a single-layer suspension of regular (uncoated) particles heated using a laser-manipulation method \cite{Nosenko:2006}.
}
\end{figure}
% ls-14.

When injected in plasma, the Janus particles formed a round single-layer suspension with a diameter of approximately $40$~mm. The suspension consisted of around $780$ particles. Due to their self-propulsion, they acquired high speeds greatly exceeding the thermal speed, which prevented the particles from forming a regular lattice \cite{Nosenko:2022}. Their trajectories, however, remained in the horizontal plane due to the strong vertical confinement, see Fig.~\ref{Fig1}(c). No out-of-plane motion of the particles was detected by the side-view camera, which rules our the mode-coupling instability (MCI) in the particle suspension \cite{Couedel:2010}. The self-propulsion force was balanced by the neutral gas drag force \cite{Epstein:1924,Liu:2003}. Similar but uncoated particles were experimentally shown to interact with each other via a screened-Coulomb potential which is approximated reasonably well by the Yukawa potential \cite{Konopka:2000,Kompaneets_PhD}.

The particle suspension was in a highly disordered (gas-like) state, as evidenced by the nearly flat pair correlation function $g(r)$, see Fig.~\ref{Fig1}(b). From the first peak of $g(r)$, the mean interparticle spacing was calculated as $a=1.06$~mm. The mean kinetic energy of the particles was $\langle E_k \rangle\approx 66$~eV (with a slight anisotropy of $1.9$\%), which is a factor of $4$ higher than the maximum $\langle E_k \rangle$ achieved in experiments with a single-layer suspension of regular (uncoated) particles heated using a laser-manipulation method \cite{Nosenko:2006}.
% $T_x=755211$~K (65.10 eV), $T_y=769699$~K (66.35 eV) -> 1.9\% anisotropy.

To characterize how chaotic the Janus particles’ motion is, we utilised a tool widely used in the field of turbulence---the longitudinal structure functions $S_p(r)$. They are two-point single-time Eulerian correlation functions defined as:

\begin{equation}\label{Sn}
S_p(r)=\left\langle\left|[{\bf v}_i(t)-{\bf v}_j(t)]\cdot {\bf r}_{ij}/r_{ij}\right|^p\right\rangle,
\end{equation}
where $r$ is the spatial scale length, ${\bf v}_i$ and ${\bf r}_i$ are respectively the velocity and position of $i$-th particle, ${\bf r}_{ij}={\bf r}_i-{\bf r}_j$, $p$ is the order of the structure function, and the averaging is performed over all pairs $(i,j)$ of particles separated by $r$. To calculate $S_p(r)$, we used the following method. In every frame of the experimental video, all possible pairs of particles within the distance of $15$~mm from the cloud center were considered (out of the total cloud radius of $\approx19$~mm) and their interparticle separations were assigned to bins of distance. The bin size was $5$ pixels, which corresponds to $0.1168$~mm. $256$ bins were considered, to the total length of $29.784$~mm, which is smaller than the maximum possible interparticle separation of $30$~mm. The averaging according to Eq.~(\ref{Sn}) ran over all such particle pairs in all $2726$ frames of the recorded video. It is important to note that the middle spatial bins accounted for $\simeq5\times10^6$ particle pairs and the highest spatial bin (corresponding to the distance of $29.784$~mm) for $7344$, which provided reasonable statistics. This procedure served to mitigate the boundary effects, as the particles at the cloud periphery were excluded from the calculations.

The significance of $S_p(r)$ in the field of turbulence is grounded in the prediction by Kolmogorov's theory of turbulence of the power-law scaling $S_p(r)\propto r^{\xi_p}$ with $\xi_p=p/3$ in the inertial range of homogeneous isotropic turbulence \cite{Landau_v6}.

The longitudinal structure functions $S_p(r)$ of the orders $p=2-6$ calculated for the suspension of Janus particles are shown in Fig.~\ref{Fig2}(a). They reach shallow minima in the vicinity of the mean interparticle spacing $a=1.06$~mm and increase for larger and especially smaller $r$. Since $S_p(r)$ are a measure of dissimilarity of the particle (radial) motion, this means that the neighboring particles have on average more similarity in their movements than more distant or closer particles. When the particles come close together, they strongly repel each other, which leads to higher values of $S_p(r)$. Note that the $S_p(r)$ functions show no clear power-law scaling, in particular, at no $r$ do they have slopes corresponding to the Kolmogorovian values of $\xi_p=p/3$ (indicated by the dotted lines).

The ESS concept is based on the observation that $\xi_3=1$ where the Kolmogorov scaling is present and hence $S_3(r)$ can be used as a ``proxy'' for the distance $r$. Therefore, the structure functions expressed as $S_p(S_3)$ rather than $S_p(r)$ should have the same scaling exponents $\xi_p=p/3$. $S_p(S_3)$ for the Janus particles are shown in Fig.~\ref{Fig2}(b). Unlike in $S_p(r)$, there is a clear power-law scaling in $S_p(S_3)$, in particular for the ``macroscopic'' length scales $r\geq a$ (shown by the solid symbols). This reveals the presence of ESS. The scaling of $S_p(S_3)$ is described surprisingly well by the Kolmogorovian exponents $\xi_p=p/3$ (shown by the dotted lines). Fitting $S_p(S_3)$ to the power law in the whole range of $r$ results in the scaling exponents that are close to $\xi_p=p/3$ for $p=2-4$ and are smaller than $p/3$ for larger $p$, see Fig.~\ref{Fig2}(c). The deviation of measured exponents $\xi_p$ from $p/3$ points at the intermittency (multiscaling) in the particle dynamics \cite{Nikora:2001}.

It is instructive to compare the scaling behavior of the Janus particle suspension to the laser-heated suspension of regular (uncoated) particles in similar experimental conditions. In the experiment of Ref.~\cite{Nosenko:2006}, laser manipulation was used to apply brief intense random kicks to the particles initially forming a regular lattice (plasma crystal). The kinetic temperature of the particles increased with the laser power applied, and above a threshold a melting transition was observed. The liquid state of the laser-heated particle suspension had properties that approximated those of a system in thermal equilibrium. Here, we calculated the structure functions $S_p(r)$ from the experimental data of Ref.~\cite{Nosenko:2006}, see Fig.~\ref{Fig3}(a).

While the general shapes of $S_p(r)$ are similar in Figs.~\ref{Fig2}(a) and \ref{Fig3}(a), there is an important difference: In the case of the laser-heated particles, $S_p(r)$ have oscillatory structures in the range of $0.5~{\rm mm}<r<2~{\rm mm}$, which arise due to the excitation of compressional waves in the particle suspension. The overall slopes of the oscillatory structures resemble those defined by the Kolmogorovian exponents $\xi_p=p/3$, as indicated by the dotted lines in Fig.~\ref{Fig3}(a). When plotted against $S_3$, the $S_p(S_3)$ functions exhibit clear power-law scaling in the whole range studied, see Fig.~\ref{Fig3}(b). Therefore, ESS is also present in the laser-heated suspension of regular particles. The scaling exponents (obtained from the power-law fits, not shown here) are close to the Kolmogorovian $\xi_p=p/3$ for $p=2-4$ and are smaller than $p/3$ for $p>4$, see Fig.~\ref{Fig3}(c). ESS-derived scaling exponents of the suspensions of active Janus particles and laser-heated regular (passive) particles turn out to be similar including intermittent (multiscaling) behaviour for $p>4$, see Figs.~\ref{Fig2}(c) and \ref{Fig3}(c). This appears to be an inherent characteristic of complex plasma.

{\it Energy spectrum}---In an active flow, the pattern of energy injection is self-organized and not imposed by the external driving. The energy injection and dissipation occur at similar spatial scales of individual particles. Therefore, the energy cascade is not necessarily present. Some energy transfer across scales is however possible. If an energy cascade develops, it may be very different from the classical energy cascade in the inertial interval of high-Reynolds-number turbulence. In the Kolmogorov theory of turbulence, the universal power-law scaling exponent of $-5/3$ is predicted for the inertial interval of fully developed turbulence \cite{Landau_v6}. In an active flow, the scaling exponent may be non-universal, that is dependent on the experimental parameters \cite{Alert:2022}.

\begin{figure}[tbp!]
\centering
\includegraphics[width=0.9\columnwidth]{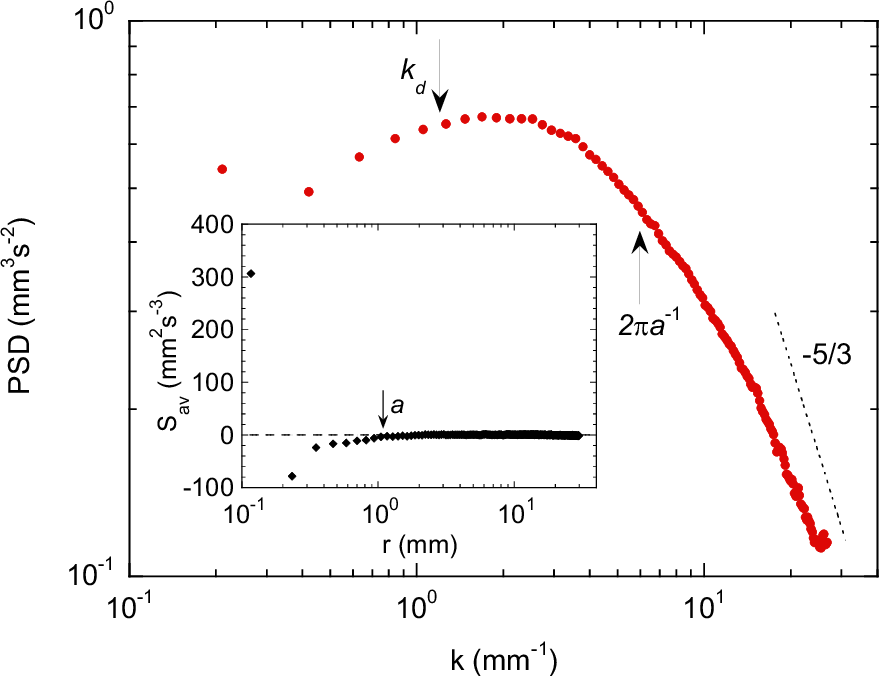}
\caption {\label {Fig4} Energy spectrum of the single-layer suspension of Janus particles. The dotted line indicates the slope of $-5/3$. The inset shows the crossed velocity-acceleration Eulerian structure function $S_{av}(r)=\langle\delta_r{\bf a}\cdot\delta_r{\bf v}\rangle$, which represents the energy flux across scales \cite{Bourgoin:2020}. The arrows indicate the mean interparticle spacing $a$, respective wave number $2\pi a^{-1}$, and the wave number $k_d$ corresponding to the inverse characteristic damping length.
}
\end{figure}
% JP-13-4-L2

The energy spectrum (power spectral density) was calculated as the Fourier transform, after subtracting the linear trend, of the Eulerian correlation function $S_2(r)$ of the Janus particles’ velocity. The energy spectrum is shown in Fig.~\ref{Fig4}. It has a broad maximum at $k\approx 2~{\rm mm}^{-1}$, which is between the wave numbers corresponding to the inverse characteristic damping length $k_d=2\pi\gamma_E/v_{\rm act}=1.2~{\rm mm}^{-1}$ (here, $\gamma_E=0.8~{\rm s}^{-1}$ is the Epstein drag rate for particles \cite{Epstein:1924,Liu:2003} and $v_{\rm act}=\sqrt{\langle E_k \rangle/m_p}=4.1$~mm/s is the mean speed of active particles \cite{footnote1}) and the inverse interparticle spacing $2\pi a^{-1}=5.9~{\rm mm}^{-1}$. This confirms that the energy input into the particle suspension occurs at the length scales related to individual particles. In the range of wave numbers $k=9-26.7~{\rm mm}^{-1}$, that is $k\gtrsim 2\pi a^{-1}$, an energy cascade develops with a scaling exponent of approximately $-1.1$, which is different from the Kolmogorov prediction of $-5/3$. This agrees well with the lack of Kolmogorovian scaling in $S_2(r)$. To find the direction of energy flow in the energy cascade, we calculated the crossed velocity-acceleration Eulerian structure function $S_{av}(r)=\langle\delta_r{\bf a}\cdot\delta_r{\bf v}\rangle$, where the brackets indicate averaging over all particle pairs and all moments of time, shown in the inset in Fig.~\ref{Fig4}. Negative values of $S_{av}(r)$ for $r\lesssim a$ indicate {\it direct} cascade (energy flows from larger to smaller scales) \cite{Bourgoin:2020}, except for the smallest $r\approx0.12$~mm, where the large positive value of $S_{av}(r)$ is due to recoiling of particles after collisions. The overall picture is then the following: Energy is injected into the system at the level of individual particles and a direct cascade of energy develops for $k\gtrsim 2\pi a^{-1}$ with a scaling exponent of $\approx-1.1$. This is different from classical two-dimensional (2D) inertial turbulence, where the energy cascade is {\it inverse} (energy flows from smaller to larger scales) \cite{Kraichnan:1967}.

In the range of wave numbers $k>2\pi a^{-1}$, the corresponding length scale is smaller than the mean interparticle distance $a$. The particles come this close to each other during their collisions. The energy dissipation in the cascade proceeds through the interparticle collisions and gas friction on individual particles \cite{Epstein:1924,Liu:2003}, by which the kinetic energy of the particles is converted into heat. Wave numbers $k>2\pi a^{-1}$ are also encountered in other situations in complex plasmas, e.g., in phonon spectra, where such $k$ are beyond the border of the first Brillouin zone \cite{Couedel:2011}.

The scaling exponent of $-1.1$ measured in this paper for a 2D system of active Janus particles is different from known universal scaling exponents reported for other active-matter systems, such as $-3/2$ for bacterial turbulence beyond a critical drive \cite{Mukherjee:2023} and $-5/3$ for a sparse assembly of interacting Marangoni surfers \cite{Bourgoin:2020}. A range of non-universal scaling exponents from $-4.5$ to $-0.3$ was reported for bacterial suspensions, cell tissues, and self-propelled colloids \cite{Alert:2022}. To find out whether the scaling exponent of our system may approach a universal exponent, e.g., $-3/2$ or $-5/3$ in certain experimental regimes, experiments with higher degree of activity are needed as suggested in Ref.~\cite{Mukherjee:2023}.

To summarize, we showed experimentally that the velocity field of a 2D system of active Janus particles manifests extended self-similarity in the whole range of distances studied. The scaling exponents are surprisingly similar to the Kolmogorov prediction for the inertial range of fully developed turbulence for the orders $p=2-4$ and are somewhat smaller for higher orders indicating intermittent particle dynamics. At the same time, the underlying longitudinal velocity structure functions of interparticle separation do not show any power-law scaling at all. The energy spectrum revealed a direct cascade, where energy flows from larger to smaller scales, for the wave numbers larger than the inverse mean interparticle spacing. Extended self-similarity was also observed in the velocity field of a 2D system of laser-heated regular (passive) particles, however, here the underlying structure functions at least display a hint at the power-law scaling near the mean interparticle distance. Therefore, ESS appears to be an inherent characteristic of complex plasmas. Note that extended self-similarity is strongly related to turbulence, but they are not equivalent to each other. Verifying whether the chaotic active flow in a 2D system of active Janus particles has other characteristics of turbulence is reserved for future research.

\acknowledgments

Philip Born is acknowledged for producing the Janus particles used in this study. Matthias Kolbe is acknowledged for taking the SEM images of Janus particles. Thomas Voigtmann is acknowledged for carefully reading the manuscript and helpful discussions.


\begin{thebibliography}{}
\bibitem{footnote1} In this estimate, the radius and mass of an uncoated particle are assumed.
\end{thebibliography}
\end{document}